\documentclass[aps,pre,twocolumn,superscriptaddress,groupedaddress]{revtex4}  
\usepackage{graphicx,epsfig}  
\usepackage{dcolumn}   
\usepackage{bm}        
\usepackage{amssymb}   
\usepackage{color}
\usepackage[caption = false]{subfig}

\usepackage{pdfcomment}
\pdfcommentsetup{final} 
 \pdfcommentsetup{color={1.0 0.0 0.0}, open=true}

\begin{document}
\title{Experimental observation of pinned solitons in a flowing dusty plasma}
\author{Garima Arora}
\email{garimagarora@gmail.com}
\author{P. Bandyopadhyay}
\author{M. G. Hariprasad}
\author{A. Sen}
\affiliation{Institute For Plasma Research, HBNI, Bhat, Gandhinagar, Gujarat, India, 382428}

\begin{abstract}
Pinned solitons are a special class of nonlinear solutions created by a  super-sonically moving object in a fluid. They move with the same velocity as the moving object and thereby remain pinned to the object. A well known hydrodynamical phenomenon, they have been shown to exist in numerical simulation studies but to date have not been observed experimentally in a plasma.  In this paper we report the first experimental excitation of pinned solitons in a dusty (complex) plasma flowing over a charged obstacle. The experiments are performed in a $\Pi$ shaped Dusty Plasma Experimental (DPEx) device in which a dusty plasma is created in the background of a DC glow discharge Ar plasma using micron sized kaolin dust particles. A biased copper wire creates a potential structure that acts as a stationary charged object over which the dust fluid is made to flow at a highly supersonic speed. Under appropriate conditions nonlinear stationary structures are observed in the laboratory frame that correspond to pinned structures moving with the speed of the obstacle in the frame of the moving fluid. A systematic study is made of the propagation characteristics of these solitons by carefully tuning the flow velocity of the dust fluid by changing the height of the potential structure. It is found that the nature of the pinned solitons changes from a single humped one to a multi-humped one and their amplitudes increase with an increase of the flow velocity of the dust fluid. The experimental findings are then qualitatively compared with the numerical solutions of a model forced Korteweg de Vries (fKdV) equation.
\end{abstract}
\maketitle
\section{Introduction}
Solitons are a well known class of stable localized nonlinear structures that have been widely observed and studied  in a variety of natural and laboratory settings including ocean waves\cite{new1990large}, excitations in optical fibres \cite{gedalin1997optical,haus1996solitons} and semiconductors \cite{barland2002cavity,stephane2002cavity}, plasmas systems\cite{zabusky1965interaction,prl_darksolitons,pintudasw,bailung_prl, samsonovsoliton}, laser plasma interactions \cite{kaw1992nonlinear,sundar2011relativistic,kumar2019excitation}, $\textit{etc}$. A number of model nonlinear evolution equations that are known to be fully integrable yield soliton solutions. The Korteweg-de Vries (KdV) equation \cite{korteweg1895xli,miura1976korteweg,ikezi1973experiments,dinkel2001soliton} is one such nonlinear partial differential equation that has been extensively employed as a model to study low frequency nonlinear wave phenomena in a plasma under conditions of weak dispersion and weak nonlinearity. The emergent  non-linear self-reinforcing wave packets that maintain their shape and identity while propagating at a constant velocity over a large distance are well represented by the exact mathematical soliton solutions of the KdV equation. While the KdV model works well for impulsive excitations of nonlinear pulses where the excitation source provides only an initial perturbation, the model is inadequate to describe experimental situations where the source continues to be operational in a continuous manner. Such is the case for example when a moving object continuously excites waves in a fluid medium. To model such a situation a KdV equation with a driving term - the so called forced KdV (fKdV) equation - has been adopted and used successfully in the past to interpret nonlinear phenomena in hydrodynamics e.g. to study nonlinear waves excited by fast moving objects in water \cite{wu1987generation,lee1989experiments,sun1985evolution,binder}.  The fKdV model yields some interesting and novel nonlinear solutions such as precursor solitons that travel ahead of the moving object at a speed faster than the object. These precursors can be excited when the speed of the moving object crosses the sound speed of the medium. The fKdV model also yields another class of soliton solutions that travel at the same speed as the moving object and remain pinned to the object as an envelope structure. These pinned solitons can be excited at a much higher speed of the object than what is required for the precursor solitons. \par
Precursor solitons in a plasma were recently observed for the first time under controlled laboratory conditions 
by flowing a dust fluid supersonically over a stationary charged object \cite{surbhiprecursor}.  In a frame where the fluid is stationary and the object is moving, the solitons were shown to propagate in the upstream direction as precursors while linear wake structures were seen to propagate in the downstream direction \cite{surbhiprecursor}. In a subsequent experiment, the propagation characteristics of these nonlinear structures were shown to depend on the shape and size of the charged object \cite{arora2019effect} over which the fluid flows. Experimental observations of this fore-wake phenomenon were well explained qualitatatively with the help of the forced KdV model equation \cite{surbhiprecursor,arora2019effect}.\par 
While propagating precursor solitons in plasmas appear to be well established both experimentally and theoretically the topic of pinned solitons has so far not received much attention. A detailed theoretical study on them over a range of amplitudes, widths and the velocities of the moving charged object was carried out by Tiwari \textit{et al.} \cite{sanat_pinned} using extensive fluid simulations. To the best of our knowledge, there has yet been no experimental study on pinned solitons in plasmas. Our present work addresses that topic and we report on the first successful excitation of such structures in a dusty plasma medium. A dusty plasma provides a convenient medium to study wave phenomena since the massive dust particles suspended in the plasma medium can be easily visualized and their low frequency behavior captured in a non-intrusive manner in a video recording. Our experiments have been carried out in the DPEx device \cite{surbhiRsi} that has been successfully used in the past to study precursor solitons \cite{surbhiprecursor,arora2019effect}. The major difference from past experimental conditions is in the speed of the dust flow which is kept highly supersonic and in the fine tuning of this speed by careful control of the height of the potential barrier created by a biased copper wire placed in the path of the dust flow. Our experiments show not only the existence of single humped pinned solitons enveloping the moving source but also multi-humped solutions as predicted in the theoretical studies of Tiwari {\it et al} \cite{sanat_pinned}. Our present findings provide the first experimental evidence of the existence of these interesting nonlinear structures and have the potential of stimulating further experimental and theoretical explorations 
in this area of nonlinear physics.
\section{Experimental Setup and procedure}
\begin{figure}[ht]
\includegraphics[scale=0.57]{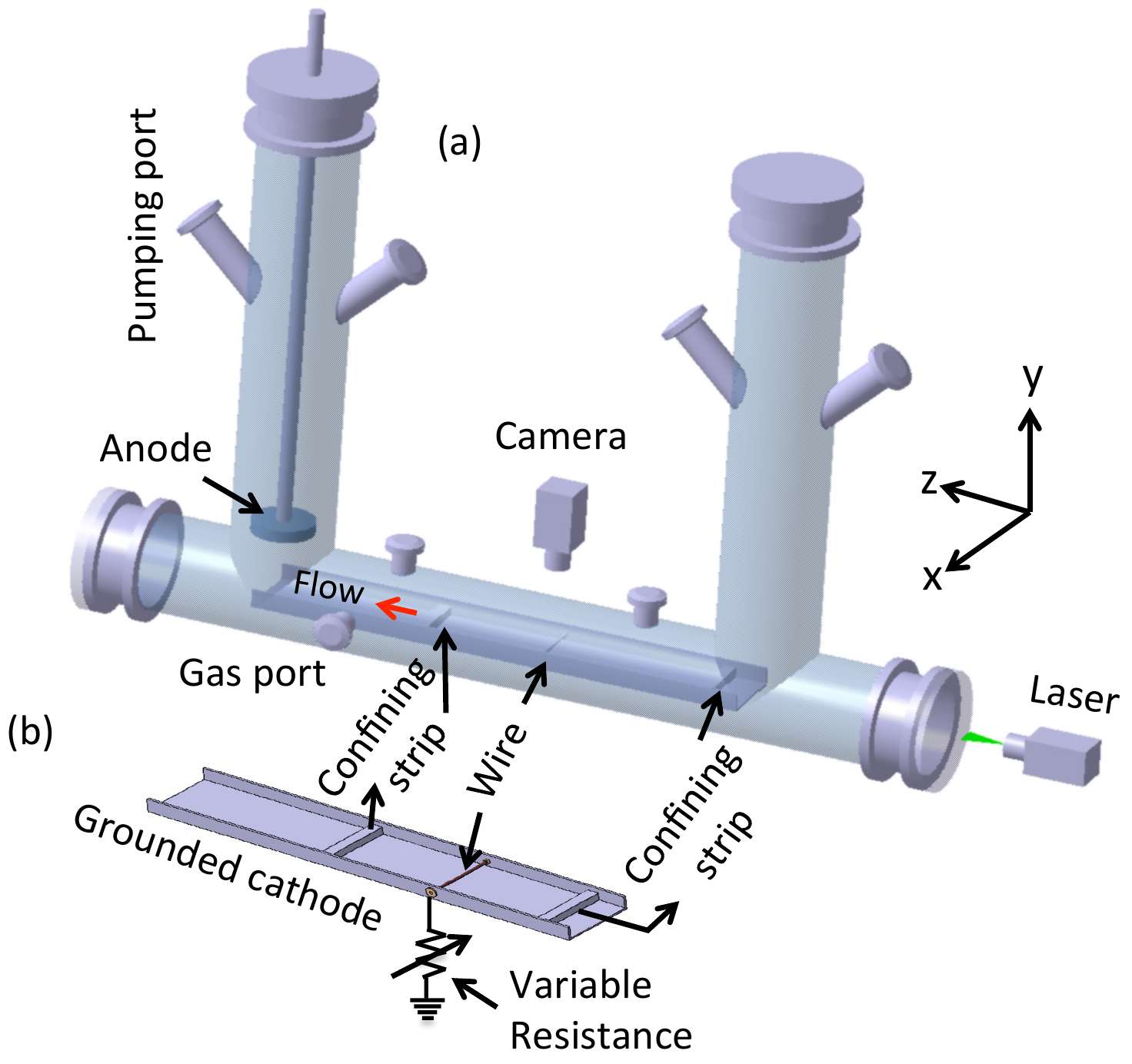}
\caption{\label{fig:fig1} A schematic diagram of the dusty plasma experimental (DPEx) setup. (b) A zoomed view of  the tray shaped grounded cathode showing the copper wire (connected with a variable resistance) and two confinement strips.}  
\end{figure}
Fig.~\ref{fig:fig1} shows the schematic diagram of the Dusty Plasma Experimental (DPEx) device in which the experiments have been carried out. It is basically an inverted $\Pi$ shaped vacuum chamber consisting of a primary cylindrical glass tube which is radially attached with two secondary tubes.  There are several radial and axial ports connected with the primary as well secondary chambers to serve various experimental needs. A disc shaped anode suspended axially from the top of the left secondary chamber and a tray type grounded cathode housed in the primary chamber are used for the production of the plasma. Two confining potential strips are placed on the cathode for confining the dust particles in the axial direction whereas the  bent sides of the cathode tray provide radial confinement. A more detailed description of the device along with its associated diagnostics are available in Ref.\cite{surbhiRsi}. A biased copper wire of diameter 1 mm and length 50 mm mounted radially on the cathode creates a potential sheath around it which acts as a stationary charged obstacle in the path of the dust flow and leads to the excitation of the pinned structures. The potential of the wire can be varied between grounded to floating  by changing a variable resistor (ranging between 10 k$\Omega$ -- 10 M$\Omega$) that is connected  in series with the wire.  The variable bias changes the height of the potential barrier and thereby the speed of the dust flow. A DC power supply (with a range of 0-1~kV and 0-500~mA) is used to strike a discharge between the electrodes. Micron sized poly-dispersive particles of diameter ranging from 2 to 5 $\mu$m are sprinkled on the cathode in between the wire and the right strip for  producing a dusty plasma. The average mass of these micron sized dust particles is estimated to be $\sim$ 8.6$\times$ $10^{-14}$ kg. \par
Initially, the chamber is pumped down to a base pressure of 0.1 Pa by fully opening the gate valve attached at the mouth of the pump. The working pressure is set to 9-15 Pa by closing the gate valve to 20 $\%$ and opening the flow meter attached to the gas port to 5-10 $\%$ as shown in Fig.~\ref{fig:fig1}. An equilibrium pressure inside the chamber is maintained through out the experiments by balancing the pumping rate and the gas flow rate. An Argon plasma is formed between the electrodes by applying a voltage in the range of $290-360$ V and the plasma parameters are measured using a single Langmuir probe and an emissive probe over a range of discharge parameters. For the present range of discharge conditions, the plasma density ($n_i$) is in the range of $\sim$ 0.5-1.5 $\times$$10^{15}$$/m^3$ and the electron temperature ($T_e$) is found to be $2-5$~eV. The above parameters are consistent with past reported measurements in similar DC glow discharge argon plasmas by several researchers \cite{fortov_pop, thompson_pop, williams_pop, pintu_prl, pramanikPLA}.  The profiles of plasma parameters over a wide range of discharge parameters are the same as presented in Ref. \cite{surbhiRsi}.\par 
 \begin{figure}[ht]
\includegraphics[scale=0.5]{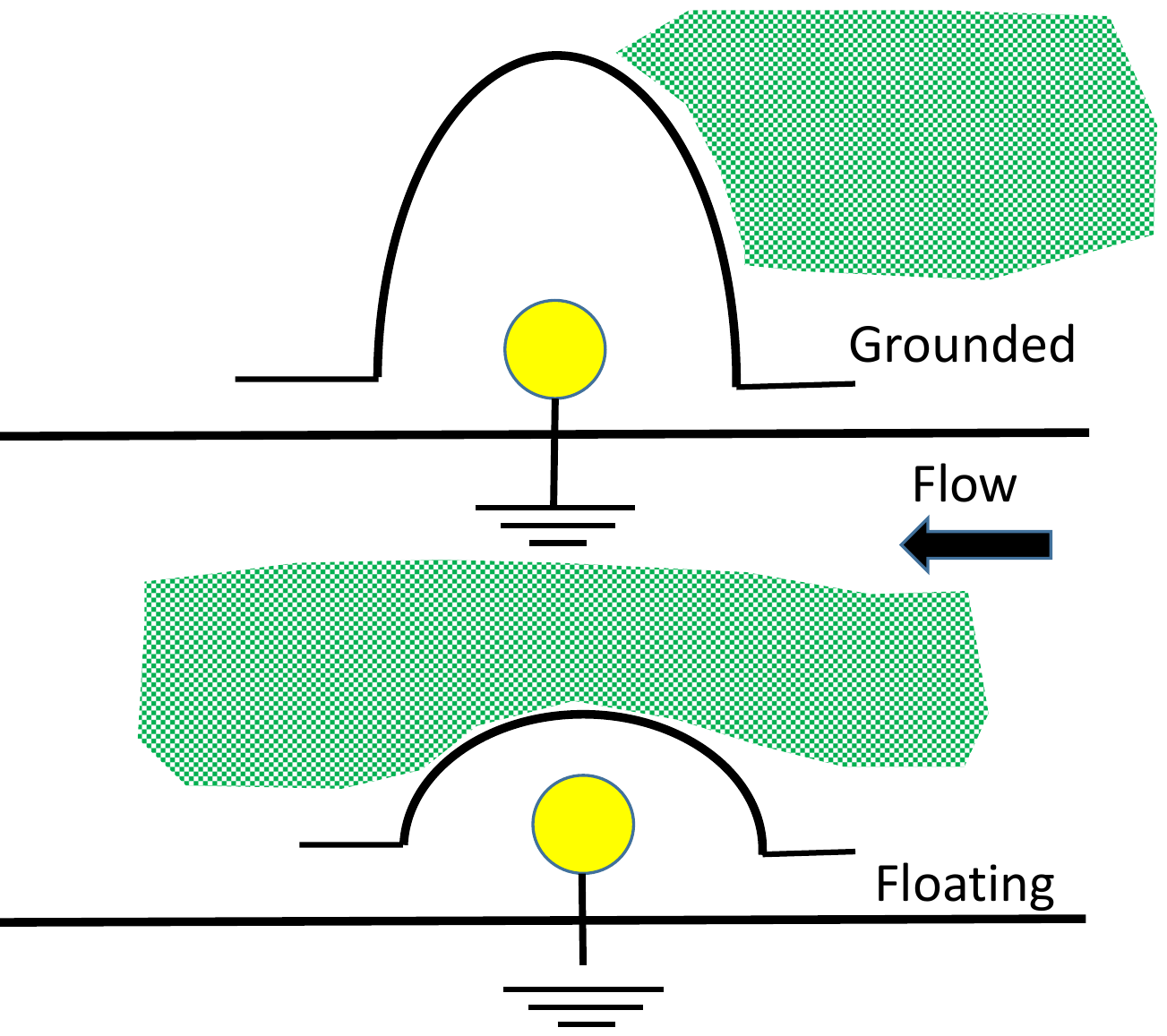}
\caption{\label{fig:fig2} (a) Cartoon sketch of the  equilibrium configuration of the dust cloud in the y-z plane prior to generating the flow. (b) Flow in the dust fluid is initiated from right to left by lowering the potential hill suddenly. The yellow circle represents the location of the copper wire shown in Fig.~1(b).}
\end{figure}
To create a dusty plasma, an equilibrium pressure $P=9$ Pa is first set and then the applied voltage is increased to 400 V so that a high electric field is created and the particles sprinkled on the cathode get charged. These particles acquire a negative charge in the plasma environment and form an equilibrium dust cloud by the balance of electrostatic and gravitational forces. In the vertical direction, the gravitational force pulls the particles in the downward direction whereas the electrostatic force (arising from the cathode sheath electric field) pushes the negatively charged particles in the upward direction. When these two forces balance each other the particle levitates in the vertical direction. The outward repulsive interactions among the negatively charged dust particles in the horizontal direction are countered by the inward repulsive force of the sheath electric field of the cathode edges in the radial direction and by that of the grounded copper wire and the right strip in the axial direction. The force balance in axial, radial, and vertical directions provide the overall confinement to create an equilibrium dust cloud. At equilibrium, the dust particles only show their thermal motion.   The levitated particles are visualised by shining a green laser light of thickness $\sim 1$~mm in the $x-z$ plane and their dynamics is captured by a fast CCD camera  looking in the same plane. The voltage is then reduced to 300 V at which a highly dense dust cloud is seen in between the grounded wire and the right confining strip. Fig.~\ref{fig:fig2}(a) shows the y-z plane of the dust fluid, which is confined in between the grounded wire and the right strip. At this specific discharge condition, the potential of the wire remains highly negative with respect to the plasma ($\sim 290$~V) and floating ($\sim 270$~V) potentials, whereas the surface potential of the dust as predicted by the CEC model \cite{khrapakcec,khrapakcec2} is $\sim -7$~V with respect to plasma potential. In other words, the potential around the grounded wire becomes negative with respect to the surface potential of the dust and as a result the particles remain confined in the potential well created by the wire and the strip.\par 
 The present set of experiments have been carried out at a discharge voltage of $V_d=300$~V and a background pressure of $9$~Pa, which are lower as compared to the past experiments on precursor solitons \cite{surbhiprecursor,arora2019effect}. For the present discharge conditions, the particles levitate at a height of ~$2.75$~cm from the cathode, which is higher than the ($\sim 2.1$~cm) height of the earlier experiments on the excitation of precursor solitons \cite{arora2019effect,surbhiprecursor}. This allows us to generate a higher supersonic flow, $M (v_f/C_{da}) \sim 1.4-3$ in the dust fluid compared to  past experimental values of $M (\sim 1.1-1.3$) \cite{arora2019effect,surbhiprecursor}.  In addition, as mentioned above, we have performed our experiments at a pressure of 9~pa which is comparatively lower than that of the previous experiments (11--12~Pa) \cite{surbhi_shock}. This leads to a lower dust neutral collision frequency of about 9~s$^{-1}$ \cite{epstein,surbhiRsi} which is half the value of 18~s$^{-1}$ reported for earlier experiments carried out on this device \cite{surbhi_shock}. We have chosen this lower dust neutral collisional regime so that dissipation becomes less important thereby avoiding  the excitation of shock waves.  The ion density and electron temperature values,  measured using a single Langmuir probe, are found to be 
 $n_i\sim$ 0.5$\times$ $10^{15}$ $/m^3$ and $T_e\sim$5 eV. The dust density ($n_d$) is approximately estimated as $\sim $ $10^{11}$ $/m^3$ from the high resolution camera images, whereas the dust temperature ($T_d$) is estimated by tracking the individual particles of the tail part of the dust cloud for 100 consecutive frames using a super Particle Identification Tracking (sPIT) \cite{fengrsi} code and comes out to be $T_d\sim$ 0.6-1.2 eV. The charge $Q_d\sim$ $10^4$e is estimated from the Collision Enhanced Plasma Collection \cite{khrapakcec,khrapakcec2} (CEC) Model for the present set of discharge conditions. With the help of plasma and dusty plasma parameters, the dust acoustic speed is theoretically estimated to be $C_{da}\sim$ 22-25 $mm/sec$. We have also conducted a separate experiment to excite dust acoustic waves (DAW) in the same experimental regime but in the absence of flow to independently determine the dust acoustic velocity. The waves were excited by applying a small electrical pulse to the wire \cite{surabhi_psst}. The average phase velocity ($C_{da}$) was obtained by tracking different crests of DAWs over time. For the range of discharge conditions of the present set of experiments, the measured phase velocity comes out to be  $C_{da} \sim 20$~mm/sec. This is in good agreement with the theoretical estimation as discussed above and the earlier measurements of Jaiswal \textit{et al.} \cite{surabhi_psst}.\par
To investigate the propagation characteristics of the nonlinear waves, a highly supersonic flow of the dust fluid ranging from 30~mm/sec to 60~mm/sec (corresponding to M=1.4 to 3)  is initiated by altering the confining potential for a particular discharge condition, p=9~Pa and V=300~V. 
The height of the potential hill is suddenly reduced from grounded potential to an intermediate potential (a potential which is less negative with respect to the particles), which generates a flow of the dust fluid from right to left as shown in Fig.~\ref{fig:fig2}(b). Within a few ms, the dust particles attain a constant velocity due to the neutral drag force \cite{garima_second, surabhi_psst} and the magnitude of this constant velocity is varied by changing the value of the intermediate potential and it is found that the maximum velocity is achieved when the wire is switched to floating potential. The flow velocity of the dust fluid is estimated using the Particle Image Velocimetry (PIV) technique \cite{thielicke2014pivlab}. It is worth mentioning that the same technique of flow generation was used by Jaiswal \textit{et al} \cite{surbhiprecursor} to excite precursor solitons in the upstream direction and wakes in downstream direction. In their experiments the range of fluid flow velocity was  $M=1.0-1.2$. In the present experiments the fluid velocity is kept higher than these earlier values and the consequent excitations of nonlinear structures studied.\par
\section{Results and Discussion}
\begin{figure}
\includegraphics[scale=0.63]{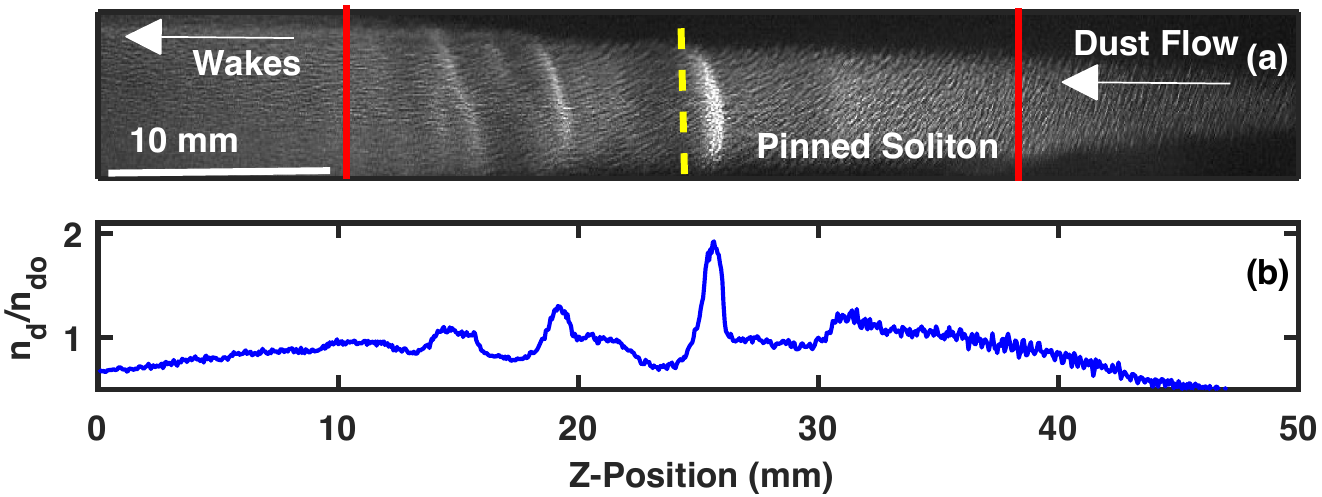}
\caption{\label{fig:fig3}(a) A typical experimental image of excitation of single pinned soliton. The dashed vertical line marks the position of the wire, whereas the distance between the two red (solid) vertical lines, placed at distances of 1.5~cms on either side of the wire,  is the approximate  spatial extent of the potential sheath. (b) Axial profile of compression factor of density perturbation extracted from Fig.~\ref{fig:fig3}(a).}
\end{figure}
The flow technique discussed above is used to generate a flow in the dust fluid with flow velocities ranging from $M\sim1.4-3$ along the axis of the chamber. The initial experiments are carried out by adjusting the resistance value in such a way that the flow velocity of the dust fluid becomes $M \sim1.5$. Fig.~\ref{fig:fig3}(a) shows a snapshot of the wave excitations occurring at this flow velocity, whereas Fig.~\ref{fig:fig3}(b) shows the axial profile of density compression, which is extracted from Fig.~\ref{fig:fig3}(a). We see a distinct one humped nonlinear structure close to the yellow dashed line which marks the location of the charged object and lower amplitude wake structures to the left of the yellow line. It is to be noted that the size of the obstacle is not the physical dimension of the wire but the width of the electrostatic sheath created by the potential of the wire. The dashed line in Fig.~\ref{fig:fig3} indicates the physical location of the wire, whereas the actual charged object  (the sheath) extends approximately 1.5 cms on both sides of the wire \cite{arora2019effect} as indicated by the red (solid) vertical lines. The distance from the wire to the single soliton in Fig.~\ref{fig:fig3} is $\sim 0.8$~mm; hence the soliton is located within the sheath and remains stationary. However,  within the experimental errors, it is hard to ascertain whether it is exactly at the center of the sheath.\par
For a quantitative analysis, the amplitude and the width of these non-linear structures are calculated by following a standard technique used in the past \cite{annibaldi2007dust,sharma2014head, pintudasw,samsonovsoliton}. The  amplitude, $A$ (defined as $\frac{n_d}{n_{do}}-1$, where $n_d$ and $n_{do}$ are the instantaneous and equilibrium dust densities, respectively) is calculated from the average pixel intensities of the images of perturbed ($I$) and equilibrium ($I_0$) dust densities using the formula $A=\frac{I}{I_{o}}-1$. It is to be noted  while calculating the amplitude, we have assumed that the light intensity captured by the camera is directly proportional to the dust density due to the linear response of the camera \cite{flanagan,merlino2012dusty}.  The width ($L$) of the pinned soliton is measured from the full width at half maximum (FWHM) of the intensity profile of perturbed dust density. The experimental data associated with the high amplitude essentially corresponds to an intense wave crest (indicative of stronger nonlinearity ) as shown in Fig.~\ref{fig:fig3}, whereas the higher width signifies broadened structures.
The amplitude ($A$), width ($L$) and the parameter ($AL^2$) for a typical single humped nonlinear structure,  measured over a number of frames for which it remained stationary, was  found to be  $A~\sim1.3310\pm 0.0649$, $L\sim 0.8575 \pm 0.0765$ and $AL^2\sim 0.9859\pm 0.0179$. As is well known, the electrostatic single humped soliton is formed when there is an exact balance of the nonlinear steepening of a wave with the broadening associated with dispersion.   As shown theoretically \cite{shukla2012nonlinear,bandyopadhyay2010effect} and experimentally \cite{sharma2014head, pintudasw,surbhiprecursor} in the past, solitons with larger amplitude always propagate with higher velocity and smaller width in such a manner that the product of the amplitude and the square of the width remains constant. 
To test this conservation property, a slightly different set of experiments was carried out by lowering the fluid flow velocity to $M \sim1.4$ but keeping the plasma parameters to be the same. The amplitude and width of the consequent excited single peaked solitary structure were found to be $A=1.2721\pm 0.0265$ and $L= 0.8645 \pm 0.0728$ respectively giving a value of $AL^2 \sim 1.0997\pm 0.01581$, which is very close to the value obtained in the earlier set of experiments. The constancy of $AL^2$ provides further evidence that such a structure is indeed a soliton  \cite{pintudasw,samsonovsoliton} - in this case a pinned soliton. \par
\begin{figure}[ht]
\includegraphics[scale=0.5]{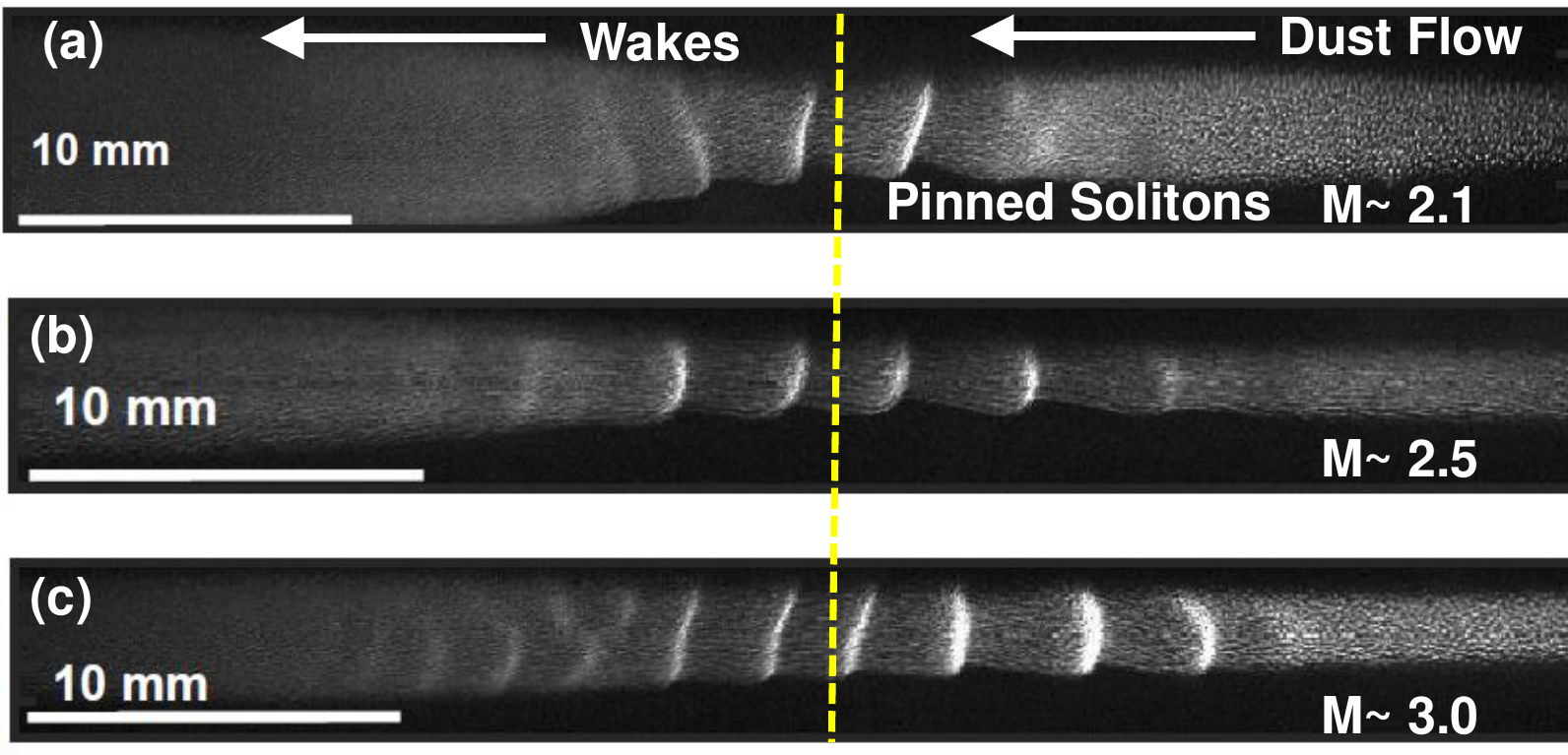}
\caption{\label{fig:fig5} Typical image of excitation of (a) double (b) four and (c) Many pinned solitons along with the wakes. The yellow dashed line represents the location of the charged object. }
\end{figure}
\begin{figure}
\includegraphics[scale=0.32]{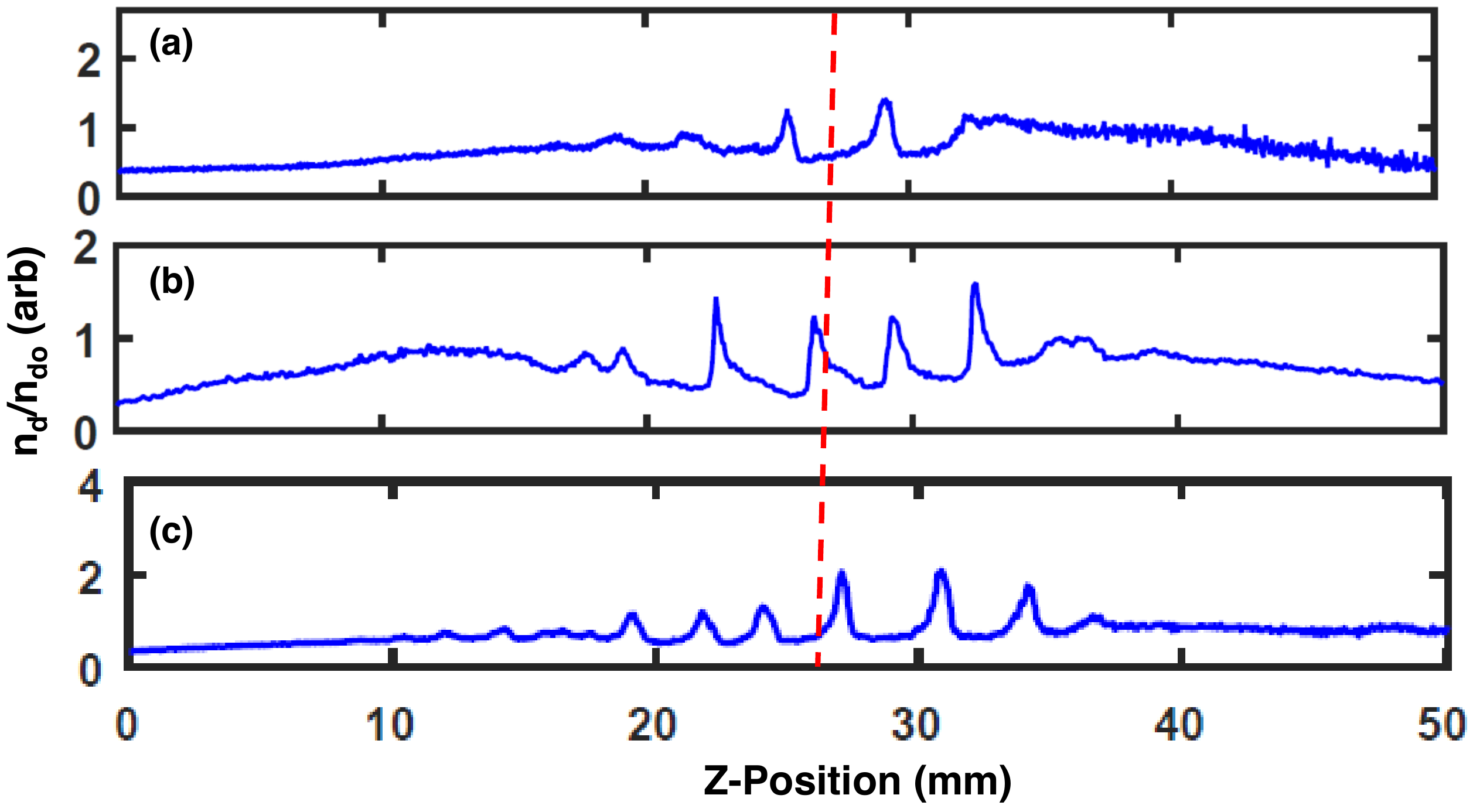}
\caption{\label{fig:fig6} Density compression factor of (a) two, (b) four and (c) many sharp peaks extracted from Fig.~\ref{fig:fig5}(a), (b) and (c). The dashed line represents the location of the wire.}
\end{figure}

When the flow velocity of the fluid is increased further to $M=2.1$ and above, by changing the wire potential under the same discharge conditions, we find a significant change in the shape of the pinned soliton.  As shown in Fig.~\ref{fig:fig5} a variety of multi-humped pinned structures appear for flow velocities M=2.1, 2.5 and 3.0 respectively.  Figure \ref{fig:fig5}(a)-(c) clearly shows that the number of humps increases with an increase in the flow velocity. The corresponding intensity profiles of the structures of Fig.~\ref{fig:fig5}(a)-(c) are plotted in  Fig.~\ref{fig:fig6}(a)-(c). Fig.~\ref{fig:fig6}(a) shows the profile of a fully developed two humped pinned soliton that remains stationary in the laboratory frame and Fig.~\ref{fig:fig6}(b) shows the profile for a similar four peaked pinned soliton.  It is to be noted that the two and four peaked structures have a symmetric profile around the wire. Interestingly, in the case of $M\sim3.0$ (see Fig.~\ref{fig:fig6}(c)), the multi-humped pinned soliton has an asymmetric structure with respect to the wire with the peaks to the right having a higher amplitude compared to the ones at the left. The characteristics of the wakes however remain the same with the increase of flow velocity and are always found to propagate in the direction of flow or in the downstream direction in the frame of the fluid.\par
\begin{figure}[ht]
\includegraphics[scale=0.43]{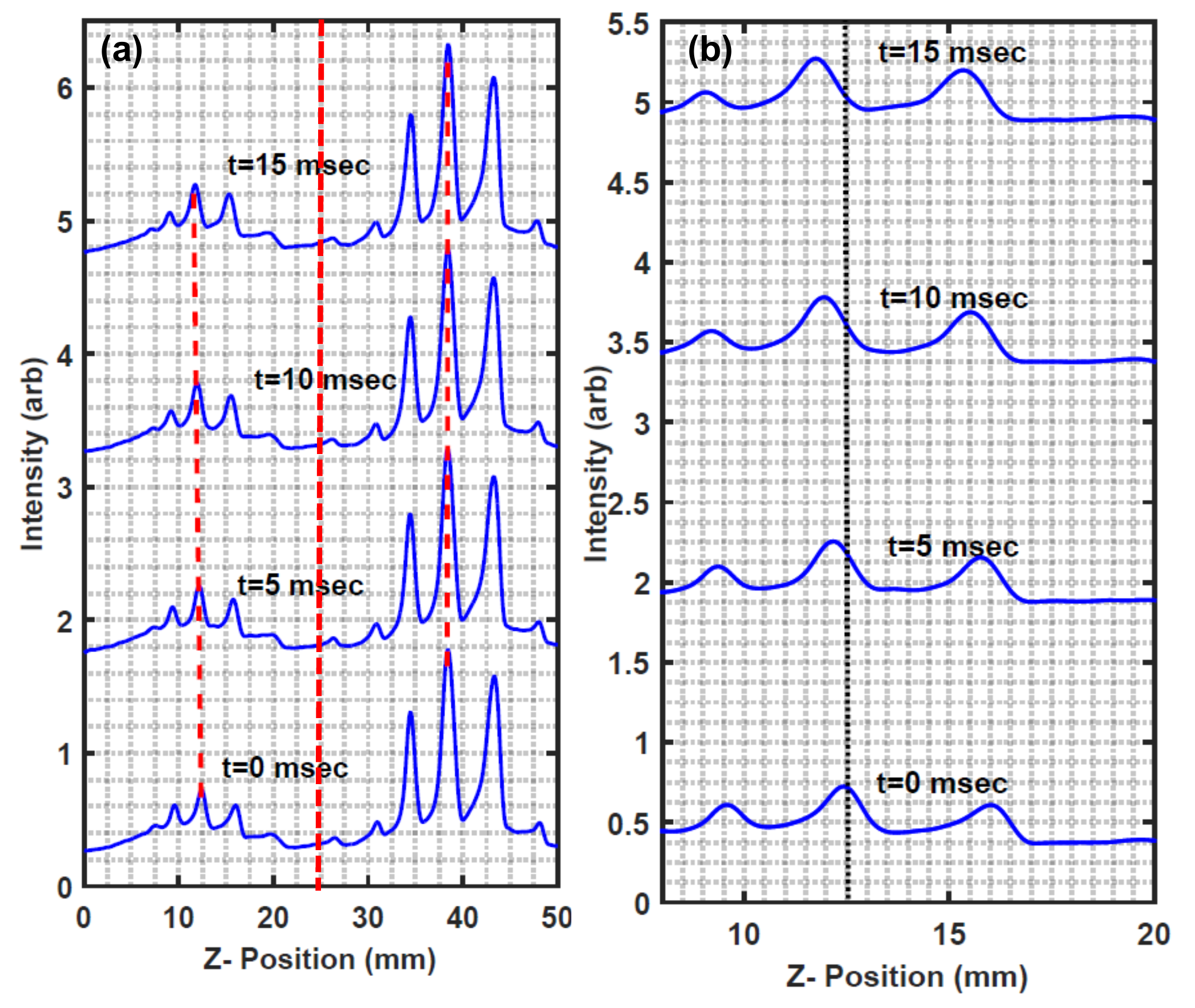}
\caption{\label{fig:fig4}(a) Intensity profile of a three humped pinned soliton and wakes over time. The dashed lines show that the higher amplitude soliton remains stationary in the laboratory frame, whereas the wakes move along the flow. The dotted line represents the location of the wire. (b) The zoomed view of a portion of Fig.~\ref{fig:fig4}(a)  clearly shows that the wakes move from right to left.}
\end{figure}

The time evolution of a three peaked soliton (as captured from subsequent frames) for the case of $M=2.3$ is plotted in Fig.~\ref{fig:fig4}(a) where the high amplitude structures in the figure represent the three peaked soliton whereas the lower amplitude structures on the left are wakes. Fig.~\ref{fig:fig4}(a) also shows that the soliton remains nearly stationary in its position in the laboratory frame of reference like the stationary charged object. In the frame of the fluid, the charged object moves from left to right along with the pinned soliton with the same velocity. The wake structures however are not stationary and propagate in the downstream direction as has been observed earlier in the experiments of Jaiswal \textit{et al.} \cite{surbhiprecursor} and also shown in Fig.~\ref{fig:fig4}(b), which is a zoomed view of Fig.~\ref{fig:fig4}(a) with only the wakes. To summarize, the highly supersonic fluid flow over a stationary charged object excites non-linear pinned solitons in the upstream direction which maintain their shape and size in the course of time, whereas the wakes propagate in the downstream direction.\par     
 \begin{figure}[ht]
\includegraphics[scale=0.8]{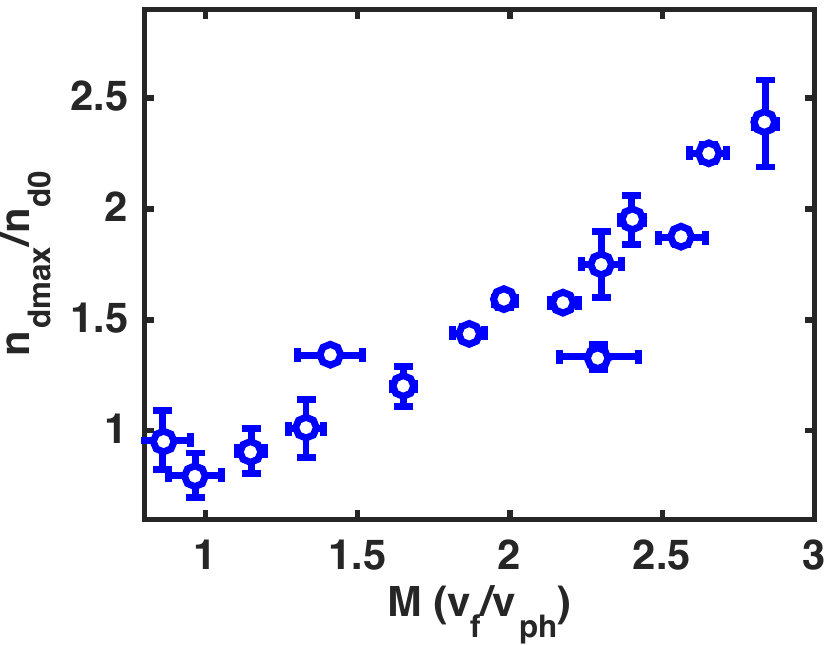}
\caption{\label{fig:fig7} Variation of maximum density compression $n_{dmax}/n_{d0}$ with the normalized flow velocity $M$.}
\end{figure}
 \begin{figure*}
\includegraphics[scale=0.55]{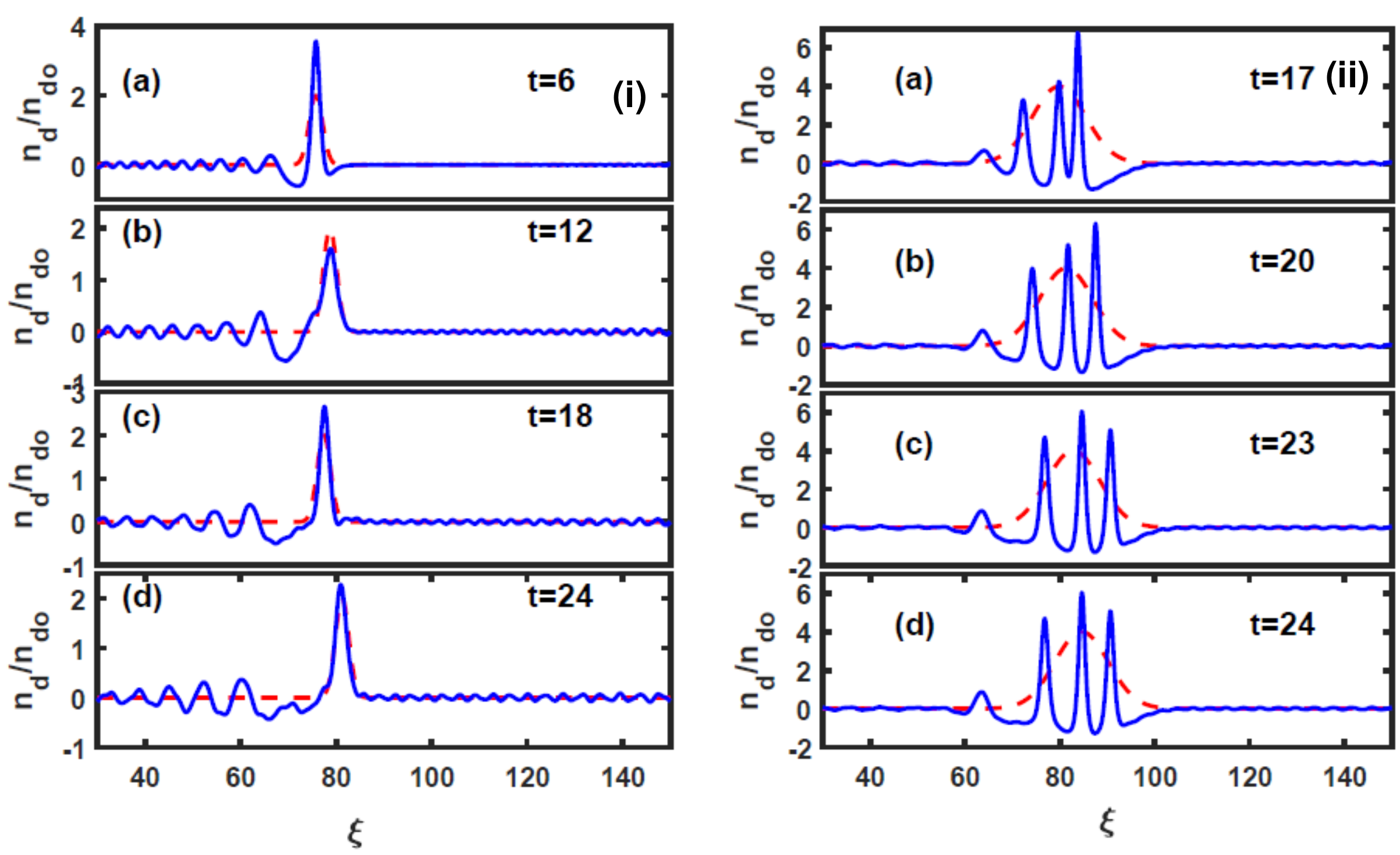}
\caption{\label{fig:fig8} Time evolution of (i) single  ($B=2.0$, $G=2.0$ and $v_d=1.5$.) and (ii) three ($B=4$, $G=8$ and $v_d=2$) pinned solitons obtained numerically by solving the f-KdV equation with $\alpha=2.0$. Solid lines represents the solitonic structure whereas the dashed line represents the source functions. The time is normalized by the inverse of dust plasma frequency.}
\end{figure*}
To quantify the dependence of the amplitude on the flow velocity the maximum values of the amplitudes of the solitons excited in the range of  M=1.5 to 3 are plotted against the Mach number in Fig.~\ref{fig:fig7}. It can be seen that with an increase in $M$, the maximum amplitude of the excited stationary structures increases. This is in qualitative agreement with the scaling observed in earlier fluid simulation studies of pinned solitons \cite{sanat_pinned}.\par
\section{Comparison with model f-KdV equation}
For a further qualitative understanding of the experimental results we have solved a model f-KdV equation numerically with a Gaussian source term \cite{SEN2015429}  and compared the numerical results with our experimental observations. The f-KdV equation for a dusty plasma is given by \cite{surbhiprecursor}, 
\begin{eqnarray}
\frac{\partial n_{d1}}{\partial t}+\alpha n_{d1}\frac{\partial n_{d1}}{\partial \xi}+\frac{1}{2}\frac{\partial^3 n_{d1}}{\partial \xi^3}=\frac{1}{2}\frac{\partial S_2}{\partial \xi},
\label{eq:eq1}
\end{eqnarray}
where $S_2$ represents the source term. $n_{d1}$ is the perturbed dust density normalized to the equilibrium density $n_{do}$ and $\xi=(z-u_{ph}t)$ is the coordinate in the wave frame moving at phase velocity $u_{ph}$ normalized to the dust acoustic speed.  The spatial coordinate $z$ is normalized by the dust Debye length ($\lambda_D$) whereas time (t) is normalized by the inverse of dust plasma frequency ($\omega_{pd}$).The coefficient $\alpha=[\delta^2+(3\delta+\sigma_i)\sigma_i+\frac{1}{2}\delta(1+\sigma_i^2)]/(\delta-1)^2$, where $\delta$ and $\sigma$ are the ratio of ion to electron density and temperature, respectively.  For our numerical investigations we have used the value of $\alpha$ to be $2$ as per the plasma parameters. A Gaussian source function is used in the form $S_2(\xi+Ft)=Bexp(-(\xi+Ft)/G)^2)$ where, $B$ and $G$ are the amplitude and width of the source function respectively and  $F=1-v_d$, with  $v_d$ the velocity of the source function with respect to the frame of fluid. The coefficient $\alpha$ in the unforced part of the model equation is associated with the nonlinear contribution arising from the plasma properties. It is the balance of this contribution with the dispersive term that produces the standard KdV soliton. As a result, the nature of the KdV soliton does depend on the value of $\alpha$. However in our case since we are not altering the plasma conditions this dependence remains unchanged. For our driven experiment we are interested in the influence of the driving amplitude ($B$) and width ($G$) of the source term as well as its velocity on the nature of the pinned solitons. In general all these three quantities influence the propagation properties of the emitted soliton. An exact analytic form of such a dependence is not known except for very special forms of the driving term (e.g. a $sech^2$ driving term gives rise to a $sech^2$ soliton as discussed in Ref. \cite{SEN2015429}). For the Gaussian shaped source we have numerically explored various values of the amplitude, width and velocity to investigate the excitation of single peaked and multi-peaked pinned solitons. One guiding factor is that the area of the Gaussian should be sufficient to produce a soliton. The additional amount is radiated away.\par
In our experiments the sheath around the wire plays the role of the charged object $S_2$. This wire was also used for the successful excitation of precursor solitons in our earlier experiments \cite{surbhiprecursor, arora2019effect}. The knowledge of the shape and size of the sheath potential from our earlier experimental measurements \cite{garima_first} indicates that  it is close to a Gaussian shape. This has prompted our choice for the form of the source function $S_2$ for the numerical solution of the f-KdV equation. The pinned solitons are numerically obtained by solving Eq.~\ref{eq:eq1} for a Gaussian source with  $B=2.0$, $G=2.0$ that is made to move with $v_d=1.5$ in accordance with the experimental condition of $M=1.5$. \par
The snap shots of $S_2$ and the normalized perturbed dust density $n_d/n_{do}$ are shown in the left panel of Fig.~\ref{fig:fig8}(a)-(d)  for $t= 6,12,18,24$, respectively. The red dashed lines in Fig.~\ref{fig:fig8} represents the position of the charged object that moves with supersonic velocity in the fluid frame. The blue solid lines represent the numerical solution of the fKdV. As can be seen, similar to the experimental results (see Figure \ref{fig:fig3}(b)), a single peaked structure representing a single humped pinned soliton is created near the source and remains stationary with respect to it. The structures to the left of the object consisting of wakes are seen to travel to the left. The slightly oscillating - decaying behaviour of the excitation observed  in the left panel Fig.~8  is a consequence of the structure trying to settle down to a soliton state by radiating waves. The right panel of Fig.~8 (a)-(d) shows snapshots of a three peaked structure for $t=$17, 20, 23, 24, respectively  as it settles down to a quasi-stationary state for a Gaussian source with  $B=4$, $G=8$ and $v_d=2.0$. We have also numerically investigated the conservation property ($AL^2$)  by looking at the single peak structure solutions of different amplitudes in the forced KdV model. For this investigation, the values of $v_d$ were changed between 1.1 to 1.5 (see Table-\ref{tab1}) for a given values of $B=2$ and $G=2$.  Table-\ref{tab1} shows that over a range of the flow velocity ($v_d$), the solitons get excited with different values of amplitudes (A) and widths (L) but the value of  the solitonic parameter $AL^2$ remains nearly constant as also seen in the case of experiments.\par 
\begin{center}
\vspace*{-.4in}
\begin{table}[h]
\caption{Solitonic parameters for different $v_d$ values}
\label{tab1}
\begin{tabular}{ |c|c|c|c|c||c|c|c| } 
\hline 
$v_d$ & Amplitude (A) & Width (L) & $AL^2$ \\
\hline \hline
{1.1}&1.354 $\pm$ 0.028 &3.18 $\pm$ 0.05 &13.67 $\pm$ 0.71  \\ 
\hline
{1.2}&1.512 $\pm$ 0.045 &3.09 $\pm$ 0.06 &14.48 $\pm$ 0.99 \\ 
\hline
{1.3}&1.816 $\pm$ 0.027 &2.71 $\pm$ 0.05 &13.35 $\pm$ 0.69 \\ 
\hline
{1.4}&2.220 $\pm$ 0.020 &2.38 $\pm$ 0.04 &12.55 $\pm$ 0.54\\ 
\hline
{1.5}&2.684 $\pm$ 0.020 &2.16 $\pm$ 0.03 &12.44 $\pm$ 0.44\\ 
\hline \hline
\end{tabular}
\end{table}
\end{center}
\vspace*{-0.2in}
 For further comparisons with the experiment, Eq.~\ref{eq:eq1} is solved for different values of the amplitude, width and speed $v_d$ of the source term to investigate the dependence of the nature of the pinned solitons on the source parameters. 
 \begin{figure}[ht]
\includegraphics[scale=0.6]{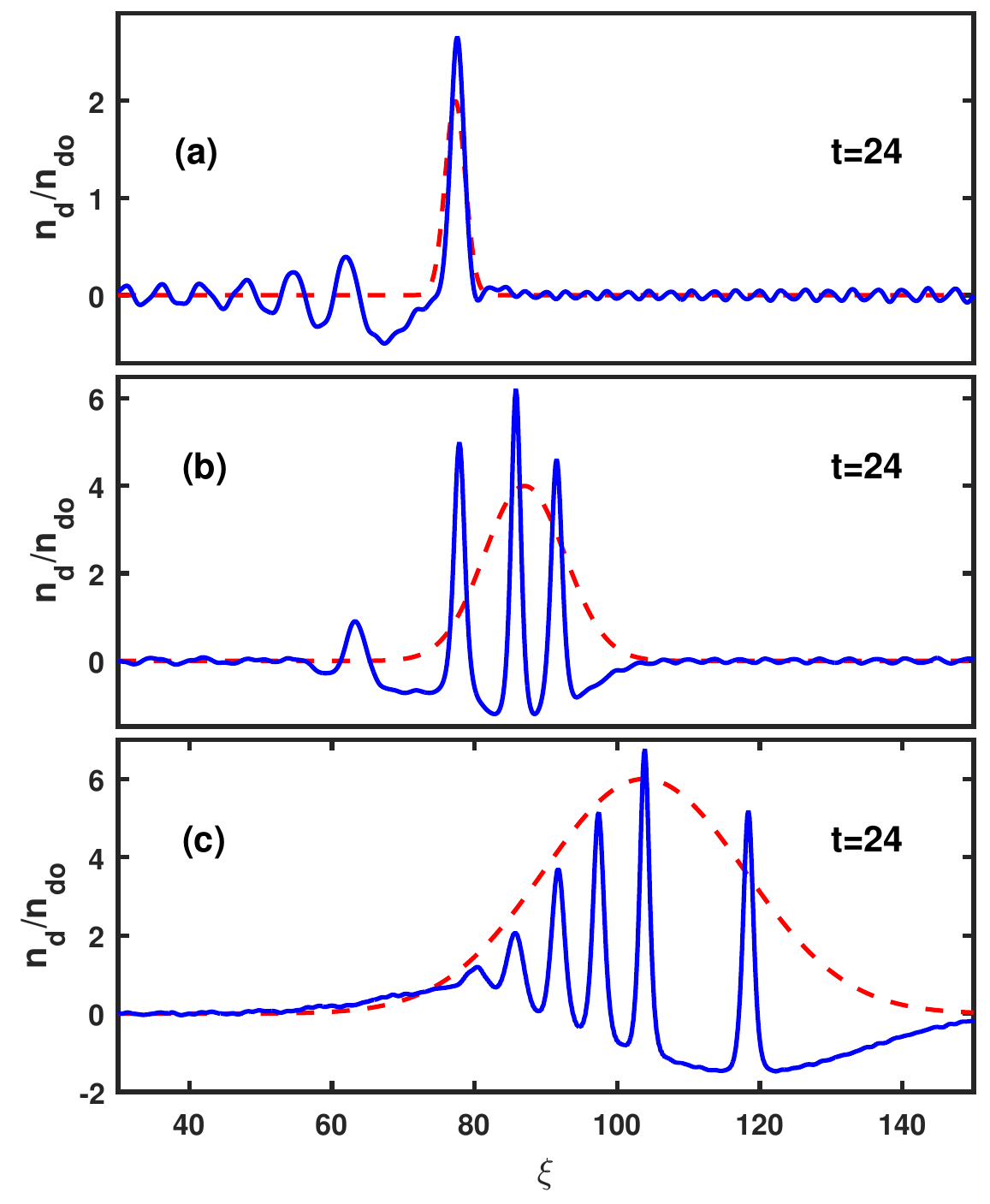}
\caption{\label{fig:fig9}Time evolution of (a) single  ($B=2.0$, $G=2.0$ and $v_d=1.5$), (b) three ($B=4$, $G=8$ and $v_d=2.0$) and (c) multiple ($B=6$, $G=20$ and $v_d=2.2$) pinned solitons obtained  by numerically solving the f-KdV equation. Solid lines represent the solitonic structures whereas the dashed lines represents the source function. Time is normalized by the inverse of dust plasma frequency.}
\end{figure}
 In the experiments the change in the velocity of the source is brought about by a change in the amplitude and width of the source e.g. the velocity of the source function was increased by increasing its amplitude which in turn increased the width. Keeping that in mind we have changed the amplitude, width and the velocity $v_d$ in the source and then solved the f-KdV equation. The results are shown in Fig~\ref{fig:fig9}(a-c) for source parameters with $B=2, 4, 6$, $G=2, 8, 20$ and $v_d=1.5,2.0, 2.2$ respectively at $t=24$. The solid lines correspond to the stationary structures excited by the moving source  and the dashed line indicates the position of the source. Similar to the experiments, it can be seen that with an increase in the velocity ($v_d$) of the source function, the number of peaks of the excited pinned solitons increases as does the value of the maximum amplitude. This is once again in  qualitative agreement with the fluid simulation results of  \cite{sanat_pinned} where it was found that the amplitude as well as the number of modulation peaks of the excited pinned solitons increased with an increase in the source velocity and amplitude. The scaling of the density compression with velocity from our numerical solutions is plotted in Fig.~\ref{fig:fig10}. Similar to the experimental findings (see Fig.~\ref{fig:fig7}), the density compression increases almost linearly with the increase of flow velocity.\par 
\begin{figure}[ht]
\includegraphics[scale=0.9]{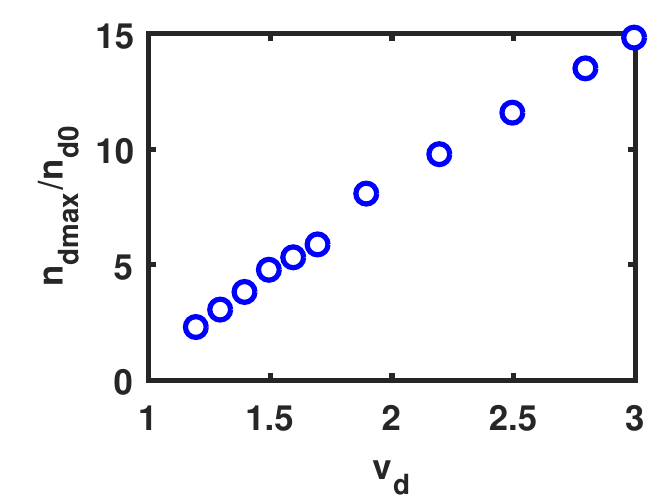}
\caption{\label{fig:fig10} Variation of maximum density compression $n_{dmax}/n_{d0}$ obtained from numerical solution with the velocity of source function  $v_d$.}
\end{figure}
\section{Conclusions}
To conclude, experimental observations of a new class of stationary solitons, known as pinned solitons, is reported in a flowing complex plasma. These experiments have been performed in the Dusty Plasma Experimental Device in which a large dust cloud was created in a DC glow discharge Argon plasma. For the purpose of generating a highly supersonic flow in the dust fluid, the discharge voltage and the background pressure were set such that the dust particles levitated at a height far above the cathode. The flow in the dust fluid was generated over the stationary charged object by changing the height of the potential hill used for the axial confinement. The highly supersonic flow of dust fluid was seen to generate multiple nonlinear stationary structures in the vicinity of the wire. In the frame  of the moving fluid, these structures  remain attached to the moving object (wire) and constitute propagating  pinned solitons. The maximum amplitude of the solitonic structures was found to display an increase with the increase of the fluid flow velocity. In addition, the number of amplitude modulations (peaks) in the density perturbation was also found to increase with the increase in the dust flow velocity. The results are in qualitative agreement with numerical solutions of the forced-KdV equation with a Gaussian source function moving with a supersonic velocity. It must be remarked here that the f-KdV model as well as the previous fluid simulations \cite{sanat_pinned} are based on a one dimensional approximation of the dynamics and hence have inherent limitations. They ignore, for example, effects associated with finite length effects in the direction transverse to the direction of propagation of the soliton that can lead to bending or curving of the soliton profiles that would show up in an experiment carried out in a finite device. Hence the model results can only serve as a qualitative guide for the interpretation of the results. However, as we see in the present experiment and as has also been observed in past experimental studies of solitons in laboratory plasma devices \cite{pintu_prl,surbhiprecursor,arora2019effect}, a KdV based model does capture the essential features of soliton excitation and its propagation characteristics.  In the present case we are able to qualitatively establish the predominant features of the pinned solitons by their stationary nature, the constancy of the quantity $AL^2$ for the one-humped structure and the existence of multi-humped forms predicted by the model. So it appears that higher dimensional effects may not be important for our experiment.
 Our present results apart from being of fundamental importance may have some potential applications in real life situations. The experimental conditions in the Earth's bow shock region with a supersonic solar wind impinging on a charged Earth is similar in configuration to our experimental set up. Likewise in the Earth's ionosphere objects like space craft and space debris get naturally charged from the ionospheric plasma and can have similar associated nonlinear excitations. The possibility of such excitations have already been discussed in the literature \cite{SEN2015429} and recognized for their utility \cite{Truitt2020}. It is hoped that such potential applications as well as the intrinsic importance of this nonlinear phenomenon can stimulate further explorations on this topic.
\begin{acknowledgments}
A.S. is thankful to the Indian National Science Academy (INSA) for their support under the INSA Senior Scientist Fellowship scheme and to AOARD for their research grant FA2386-18-1-4022.
\end{acknowledgments}


\begin{thebibliography}{47}
\expandafter\ifx\csname natexlab\endcsname\relax\def\natexlab#1{#1}\fi
\expandafter\ifx\csname bibnamefont\endcsname\relax
  \def\bibnamefont#1{#1}\fi
\expandafter\ifx\csname bibfnamefont\endcsname\relax
  \def\bibfnamefont#1{#1}\fi
\expandafter\ifx\csname citenamefont\endcsname\relax
  \def\citenamefont#1{#1}\fi
\expandafter\ifx\csname url\endcsname\relax
  \def\url#1{\texttt{#1}}\fi
\expandafter\ifx\csname urlprefix\endcsname\relax\def\urlprefix{URL }\fi
\providecommand{\bibinfo}[2]{#2}
\providecommand{\eprint}[2][]{\url{#2}}

\bibitem[{\citenamefont{New and Pingree}(1990)}]{new1990large}
\bibinfo{author}{\bibfnamefont{A.}~\bibnamefont{New}} \bibnamefont{and}
  \bibinfo{author}{\bibfnamefont{R.}~\bibnamefont{Pingree}},
  \bibinfo{journal}{Deep Sea Research Part A. Oceanographic Research Papers}
  \textbf{\bibinfo{volume}{37}}, \bibinfo{pages}{513} (\bibinfo{year}{1990}).

\bibitem[{\citenamefont{Gedalin et~al.}(1997)\citenamefont{Gedalin, Scott, and
  Band}}]{gedalin1997optical}
\bibinfo{author}{\bibfnamefont{M.}~\bibnamefont{Gedalin}},
  \bibinfo{author}{\bibfnamefont{T. C.}~\bibnamefont{Scott}}, \bibnamefont{and}
  \bibinfo{author}{\bibfnamefont{Y. B.}~\bibnamefont{Band}},
  \bibinfo{journal}{Physical review letters} \textbf{\bibinfo{volume}{78}},
  \bibinfo{pages}{448} (\bibinfo{year}{1997}).

\bibitem[{\citenamefont{Haus and Wong}(1996)}]{haus1996solitons}
\bibinfo{author}{\bibfnamefont{H.~A.} \bibnamefont{Haus}} \bibnamefont{and}
  \bibinfo{author}{\bibfnamefont{W.~S.} \bibnamefont{Wong}},
  \bibinfo{journal}{Reviews of modern physics} \textbf{\bibinfo{volume}{68}},
  \bibinfo{pages}{423} (\bibinfo{year}{1996}).

\bibitem[{\citenamefont{Barland et~al.}(2002)\citenamefont{Barland, Tredicce,
  Brambilla, Lugiato, Balle, Giudici, Maggipinto, Spinelli, Tissoni, Knoedl
  et~al.}}]{barland2002cavity}
\bibinfo{author}{\bibfnamefont{S.}~\bibnamefont{Barland}},
  \bibinfo{author}{\bibfnamefont{J.~R.} \bibnamefont{Tredicce}},
  \bibinfo{author}{\bibfnamefont{M.}~\bibnamefont{Brambilla}},
  \bibinfo{author}{\bibfnamefont{L.~A.} \bibnamefont{Lugiato}},
  \bibinfo{author}{\bibfnamefont{S.}~\bibnamefont{Balle}},
  \bibinfo{author}{\bibfnamefont{M.}~\bibnamefont{Giudici}},
  \bibinfo{author}{\bibfnamefont{T.}~\bibnamefont{Maggipinto}},
  \bibinfo{author}{\bibfnamefont{L.}~\bibnamefont{Spinelli}},
  \bibinfo{author}{\bibfnamefont{G.}~\bibnamefont{Tissoni}},
  \bibinfo{author}{\bibfnamefont{T.}~\bibnamefont{Knoedl}},
  \bibnamefont{et~al.}, \bibinfo{journal}{Nature}
  \textbf{\bibinfo{volume}{419}}, \bibinfo{pages}{699} (\bibinfo{year}{2002}).

\bibitem[{\citenamefont{Stephane et~al.}(2002)\citenamefont{Stephane, Tredicce,
  Massimo, Lugiato, Salvador, Massimo, Tommaso, Spinelli, Giovanna, Thomas
  et~al.}}]{stephane2002cavity}
\bibinfo{author}{\bibfnamefont{B.}~\bibnamefont{Stephane}},
  \bibinfo{author}{\bibfnamefont{J.~R.} \bibnamefont{Tredicce}},
  \bibinfo{author}{\bibfnamefont{B.}~\bibnamefont{Massimo}},
  \bibinfo{author}{\bibfnamefont{L.~A.} \bibnamefont{Lugiato}},
  \bibinfo{author}{\bibfnamefont{B.}~\bibnamefont{Salvador}},
  \bibinfo{author}{\bibfnamefont{G.}~\bibnamefont{Massimo}},
  \bibinfo{author}{\bibfnamefont{M.}~\bibnamefont{Tommaso}},
  \bibinfo{author}{\bibfnamefont{L.}~\bibnamefont{Spinelli}},
  \bibinfo{author}{\bibfnamefont{T.}~\bibnamefont{Giovanna}},
  \bibinfo{author}{\bibfnamefont{K.}~\bibnamefont{Thomas}},
  \bibnamefont{et~al.}, \bibinfo{journal}{Nature}
  \textbf{\bibinfo{volume}{419}}, \bibinfo{pages}{699} (\bibinfo{year}{2002}).

\bibitem[{\citenamefont{Zabusky and Kruskal}(1965)}]{zabusky1965interaction}
\bibinfo{author}{\bibfnamefont{N.~J.} \bibnamefont{Zabusky}} \bibnamefont{and}
  \bibinfo{author}{\bibfnamefont{M.~D.} \bibnamefont{Kruskal}},
  \bibinfo{journal}{Physical review letters} \textbf{\bibinfo{volume}{15}},
  \bibinfo{pages}{240} (\bibinfo{year}{1965}).

\bibitem[{\citenamefont{Heidemann et~al.}(2009)\citenamefont{Heidemann,
  Zhdanov, S\"utterlin, Thomas, and Morfill}}]{prl_darksolitons}
\bibinfo{author}{\bibfnamefont{R.}~\bibnamefont{Heidemann}},
  \bibinfo{author}{\bibfnamefont{S.}~\bibnamefont{Zhdanov}},
  \bibinfo{author}{\bibfnamefont{R.}~\bibnamefont{S\"utterlin}},
  \bibinfo{author}{\bibfnamefont{H.~M.} \bibnamefont{Thomas}},
  \bibnamefont{and} \bibinfo{author}{\bibfnamefont{G.~E.}
  \bibnamefont{Morfill}}, \bibinfo{journal}{Phys. Rev. Lett.}
  \textbf{\bibinfo{volume}{102}}, \bibinfo{pages}{135002}
  (\bibinfo{year}{2009}),
  \urlprefix\url{https://link.aps.org/doi/10.1103/PhysRevLett.102.135002}.

\bibitem[{\citenamefont{Bandyopadhyay
  et~al.}(2008{\natexlab{a}})\citenamefont{Bandyopadhyay, Prasad, Sen, and
  Kaw}}]{pintudasw}
\bibinfo{author}{\bibfnamefont{P.}~\bibnamefont{Bandyopadhyay}},
  \bibinfo{author}{\bibfnamefont{G.}~\bibnamefont{Prasad}},
  \bibinfo{author}{\bibfnamefont{A.}~\bibnamefont{Sen}}, \bibnamefont{and}
  \bibinfo{author}{\bibfnamefont{P.~K.} \bibnamefont{Kaw}},
  \bibinfo{journal}{Phys. Rev. Lett.} \textbf{\bibinfo{volume}{101}},
  \bibinfo{pages}{065006} (\bibinfo{year}{2008}{\natexlab{a}}),
  \urlprefix\url{https://link.aps.org/doi/10.1103/PhysRevLett.101.065006}.

\bibitem[{\citenamefont{Nakamura et~al.}(1999)\citenamefont{Nakamura, Bailung,
  and Shukla}}]{bailung_prl}
\bibinfo{author}{\bibfnamefont{Y.}~\bibnamefont{Nakamura}},
  \bibinfo{author}{\bibfnamefont{H.}~\bibnamefont{Bailung}}, \bibnamefont{and}
  \bibinfo{author}{\bibfnamefont{P.~K.} \bibnamefont{Shukla}},
  \bibinfo{journal}{Phys. Rev. Lett.} \textbf{\bibinfo{volume}{83}},
  \bibinfo{pages}{1602} (\bibinfo{year}{1999}),
  \urlprefix\url{https://link.aps.org/doi/10.1103/PhysRevLett.83.1602}.

\bibitem[{\citenamefont{Samsonov et~al.}(2002)\citenamefont{Samsonov, Ivlev,
  Quinn, Morfill, and Zhdanov}}]{samsonovsoliton}
\bibinfo{author}{\bibfnamefont{D.}~\bibnamefont{Samsonov}},
  \bibinfo{author}{\bibfnamefont{A.~V.} \bibnamefont{Ivlev}},
  \bibinfo{author}{\bibfnamefont{R.~A.} \bibnamefont{Quinn}},
  \bibinfo{author}{\bibfnamefont{G.}~\bibnamefont{Morfill}}, \bibnamefont{and}
  \bibinfo{author}{\bibfnamefont{S.}~\bibnamefont{Zhdanov}},
  \bibinfo{journal}{Phys. Rev. Lett.} \textbf{\bibinfo{volume}{88}},
  \bibinfo{pages}{095004} (\bibinfo{year}{2002}),
  \urlprefix\url{https://link.aps.org/doi/10.1103/PhysRevLett.88.095004}.

\bibitem[{\citenamefont{Kaw et~al.}(1992)\citenamefont{Kaw, Sen, and
  Katsouleas}}]{kaw1992nonlinear}
\bibinfo{author}{\bibfnamefont{P. K.}~\bibnamefont{Kaw}},
  \bibinfo{author}{\bibfnamefont{A.}~\bibnamefont{Sen}}, \bibnamefont{and}
  \bibinfo{author}{\bibfnamefont{T.}~\bibnamefont{Katsouleas}},
  \bibinfo{journal}{Physical review letters} \textbf{\bibinfo{volume}{68}},
  \bibinfo{pages}{3172} (\bibinfo{year}{1992}).

\bibitem[{\citenamefont{Sundar et~al.}(2011)\citenamefont{Sundar, Das, Saxena,
  Kaw, and Sen}}]{sundar2011relativistic}
\bibinfo{author}{\bibfnamefont{S.}~\bibnamefont{Sundar}},
  \bibinfo{author}{\bibfnamefont{A.}~\bibnamefont{Das}},
  \bibinfo{author}{\bibfnamefont{V.}~\bibnamefont{Saxena}},
  \bibinfo{author}{\bibfnamefont{P.}~\bibnamefont{Kaw}}, \bibnamefont{and}
  \bibinfo{author}{\bibfnamefont{A.}~\bibnamefont{Sen}},
  \bibinfo{journal}{Physics of Plasmas} \textbf{\bibinfo{volume}{18}},
  \bibinfo{pages}{112112} (\bibinfo{year}{2011}).

\bibitem[{\citenamefont{Kumar et~al.}(2019)\citenamefont{Kumar, Shukla, Verma,
  Das, and Kaw}}]{kumar2019excitation}
\bibinfo{author}{\bibfnamefont{A.}~\bibnamefont{Kumar}},
  \bibinfo{author}{\bibfnamefont{C.}~\bibnamefont{Shukla}},
  \bibinfo{author}{\bibfnamefont{D.}~\bibnamefont{Verma}},
  \bibinfo{author}{\bibfnamefont{A.}~\bibnamefont{Das}}, \bibnamefont{and}
  \bibinfo{author}{\bibfnamefont{P.}~\bibnamefont{Kaw}},
  \bibinfo{journal}{Plasma Physics and Controlled Fusion}
  \textbf{\bibinfo{volume}{61}}, \bibinfo{pages}{065009}
  (\bibinfo{year}{2019}).

\bibitem[{\citenamefont{Korteweg and De~Vries}(1895)}]{korteweg1895xli}
\bibinfo{author}{\bibfnamefont{D.~J.} \bibnamefont{Korteweg}} \bibnamefont{and}
  \bibinfo{author}{\bibfnamefont{G.}~\bibnamefont{De~Vries}},
  \bibinfo{journal}{The London, Edinburgh, and Dublin Philosophical Magazine
  and Journal of Science} \textbf{\bibinfo{volume}{39}}, \bibinfo{pages}{422}
  (\bibinfo{year}{1895}).

\bibitem[{\citenamefont{Miura}(1976)}]{miura1976korteweg}
\bibinfo{author}{\bibfnamefont{R.~M.} \bibnamefont{Miura}},
  \bibinfo{journal}{SIAM review} \textbf{\bibinfo{volume}{18}},
  \bibinfo{pages}{412} (\bibinfo{year}{1976}).

\bibitem[{\citenamefont{Ikezi}(1973)}]{ikezi1973experiments}
\bibinfo{author}{\bibfnamefont{H.}~\bibnamefont{Ikezi}}, \bibinfo{journal}{The
  Physics of Fluids} \textbf{\bibinfo{volume}{16}}, \bibinfo{pages}{1668}
  (\bibinfo{year}{1973}).

\bibitem[{\citenamefont{Dinkel et~al.}(2001)\citenamefont{Dinkel, Setzer,
  Rawal, and Lonngren}}]{dinkel2001soliton}
\bibinfo{author}{\bibfnamefont{J.~N.} \bibnamefont{Dinkel}},
  \bibinfo{author}{\bibfnamefont{C.}~\bibnamefont{Setzer}},
  \bibinfo{author}{\bibfnamefont{S.}~\bibnamefont{Rawal}}, \bibnamefont{and}
  \bibinfo{author}{\bibfnamefont{K.~E.} \bibnamefont{Lonngren}},
  \bibinfo{journal}{Chaos, Solitons , Fractals} \textbf{\bibinfo{volume}{12}},
  \bibinfo{pages}{91 } (\bibinfo{year}{2001}), ISSN \bibinfo{issn}{0960-0779},
  \urlprefix\url{http://www.sciencedirect.com/science/article/pii/S0960077999001733}.

\bibitem[{\citenamefont{Wu}(1987)}]{wu1987generation}
\bibinfo{author}{\bibfnamefont{T.~Y.-T.} \bibnamefont{Wu}},
  \bibinfo{journal}{Journal of fluid mechanics} \textbf{\bibinfo{volume}{184}},
  \bibinfo{pages}{75} (\bibinfo{year}{1987}).

\bibitem[{\citenamefont{Lee et~al.}(1989)\citenamefont{Lee, Yates, and
  Wu}}]{lee1989experiments}
\bibinfo{author}{\bibfnamefont{S.-J.} \bibnamefont{Lee}},
  \bibinfo{author}{\bibfnamefont{G.~T.} \bibnamefont{Yates}}, \bibnamefont{and}
  \bibinfo{author}{\bibfnamefont{T.~Y.} \bibnamefont{Wu}},
  \bibinfo{journal}{Journal of Fluid Mechanics} \textbf{\bibinfo{volume}{199}},
  \bibinfo{pages}{569} (\bibinfo{year}{1989}).

\bibitem[{\citenamefont{Sun}(1985)}]{sun1985evolution}
\bibinfo{author}{\bibfnamefont{M.}~\bibnamefont{Sun}}, \bibinfo{journal}{The
  60th Anniversary Volume Mechanics Essays} pp. \bibinfo{pages}{17--25}
  (\bibinfo{year}{1985}).

\bibitem[{\citenamefont{Binder et~al.}(2014)\citenamefont{Binder, Blyth, and
  Balasuriya}}]{binder}
\bibinfo{author}{\bibfnamefont{B.~J.} \bibnamefont{Binder}},
  \bibinfo{author}{\bibfnamefont{M.~G.} \bibnamefont{Blyth}}, \bibnamefont{and}
  \bibinfo{author}{\bibfnamefont{S.}~\bibnamefont{Balasuriya}},
  \bibinfo{journal}{EPL (Europhysics Letters)} \textbf{\bibinfo{volume}{105}},
  \bibinfo{pages}{44003} (\bibinfo{year}{2014}),
  \urlprefix\url{http://stacks.iop.org/0295-5075/105/i=4/a=44003}.

\bibitem[{\citenamefont{Jaiswal
  et~al.}(2016{\natexlab{a}})\citenamefont{Jaiswal, Bandyopadhyay, and
  Sen}}]{surbhiprecursor}
\bibinfo{author}{\bibfnamefont{S.}~\bibnamefont{Jaiswal}},
  \bibinfo{author}{\bibfnamefont{P.}~\bibnamefont{Bandyopadhyay}},
  \bibnamefont{and} \bibinfo{author}{\bibfnamefont{A.}~\bibnamefont{Sen}},
  \bibinfo{journal}{Phys. Rev. E} \textbf{\bibinfo{volume}{93}},
  \bibinfo{pages}{041201(R)} (\bibinfo{year}{2016}{\natexlab{a}}),
  \urlprefix\url{https://link.aps.org/doi/10.1103/PhysRevE.93.041201}.

\bibitem[{\citenamefont{Arora et~al.}(2019{\natexlab{a}})\citenamefont{Arora,
  Bandyopadhyay, Hariprasad, and Sen}}]{arora2019effect}
\bibinfo{author}{\bibfnamefont{G.}~\bibnamefont{Arora}},
  \bibinfo{author}{\bibfnamefont{P.}~\bibnamefont{Bandyopadhyay}},
  \bibinfo{author}{\bibfnamefont{M.~G.} \bibnamefont{Hariprasad}},
  \bibnamefont{and} \bibinfo{author}{\bibfnamefont{A.}~\bibnamefont{Sen}},
  \bibinfo{journal}{Physics of Plasmas} \textbf{\bibinfo{volume}{26}},
  \bibinfo{pages}{093701} (\bibinfo{year}{2019}{\natexlab{a}}).

\bibitem[{\citenamefont{Kumar~Tiwari and Sen}(2016)}]{sanat_pinned}
\bibinfo{author}{\bibfnamefont{S.}~\bibnamefont{Kumar~Tiwari}}
  \bibnamefont{and} \bibinfo{author}{\bibfnamefont{A.}~\bibnamefont{Sen}},
  \bibinfo{journal}{Physics of Plasmas} \textbf{\bibinfo{volume}{23}},
  \bibinfo{pages}{022301} (\bibinfo{year}{2016}),
  \eprint{https://doi.org/10.1063/1.4941092},
  \urlprefix\url{https://doi.org/10.1063/1.4941092}.

\bibitem[{\citenamefont{Jaiswal et~al.}(2015)\citenamefont{Jaiswal,
  Bandyopadhyay, and Sen}}]{surbhiRsi}
\bibinfo{author}{\bibfnamefont{S.}~\bibnamefont{Jaiswal}},
  \bibinfo{author}{\bibfnamefont{P.}~\bibnamefont{Bandyopadhyay}},
  \bibnamefont{and} \bibinfo{author}{\bibfnamefont{A.}~\bibnamefont{Sen}},
  \bibinfo{journal}{Review of Scientific Instruments}
  \textbf{\bibinfo{volume}{86}}, \bibinfo{pages}{113503}
  (\bibinfo{year}{2015}),
  \eprint{https://aip.scitation.org/doi/pdf/10.1063/1.4935608},
  \urlprefix\url{https://aip.scitation.org/doi/abs/10.1063/1.4935608}.

\bibitem[{\citenamefont{Fortov et~al.}(2000)\citenamefont{Fortov, Khrapak,
  Khrapak, Molotkov, Nefedov, Petrov, and Torchinsky}}]{fortov_pop}
\bibinfo{author}{\bibfnamefont{V.~E.} \bibnamefont{Fortov}},
  \bibinfo{author}{\bibfnamefont{A.~G.} \bibnamefont{Khrapak}},
  \bibinfo{author}{\bibfnamefont{S.~A.} \bibnamefont{Khrapak}},
  \bibinfo{author}{\bibfnamefont{V.~I.} \bibnamefont{Molotkov}},
  \bibinfo{author}{\bibfnamefont{A.~P.} \bibnamefont{Nefedov}},
  \bibinfo{author}{\bibfnamefont{O.~F.} \bibnamefont{Petrov}},
  \bibnamefont{and} \bibinfo{author}{\bibfnamefont{V.~M.}
  \bibnamefont{Torchinsky}}, \bibinfo{journal}{Physics of Plasmas}
  \textbf{\bibinfo{volume}{7}}, \bibinfo{pages}{1374} (\bibinfo{year}{2000}).

\bibitem[{\citenamefont{Thompson et~al.}(1997)\citenamefont{Thompson, Barkan,
  D’Angelo, and Merlino}}]{thompson_pop}
\bibinfo{author}{\bibfnamefont{C.}~\bibnamefont{Thompson}},
  \bibinfo{author}{\bibfnamefont{A.}~\bibnamefont{Barkan}},
  \bibinfo{author}{\bibfnamefont{N.}~\bibnamefont{D’Angelo}},
  \bibnamefont{and} \bibinfo{author}{\bibfnamefont{R.~L.}
  \bibnamefont{Merlino}}, \bibinfo{journal}{Physics of Plasmas}
  \textbf{\bibinfo{volume}{4}}, \bibinfo{pages}{2331} (\bibinfo{year}{1997}).

\bibitem[{\citenamefont{Williams et~al.}(2008)\citenamefont{Williams, Thomas,
  and Marcus}}]{williams_pop}
\bibinfo{author}{\bibfnamefont{J.~D.} \bibnamefont{Williams}},
  \bibinfo{author}{\bibfnamefont{E.}~\bibnamefont{Thomas}}, \bibnamefont{and}
  \bibinfo{author}{\bibfnamefont{L.}~\bibnamefont{Marcus}},
  \bibinfo{journal}{Physics of Plasmas} \textbf{\bibinfo{volume}{15}},
  \bibinfo{pages}{043704} (\bibinfo{year}{2008}).

\bibitem[{\citenamefont{Bandyopadhyay
  et~al.}(2008{\natexlab{b}})\citenamefont{Bandyopadhyay, Prasad, Sen, and
  Kaw}}]{pintu_prl}
\bibinfo{author}{\bibfnamefont{P.}~\bibnamefont{Bandyopadhyay}},
  \bibinfo{author}{\bibfnamefont{G.}~\bibnamefont{Prasad}},
  \bibinfo{author}{\bibfnamefont{A.}~\bibnamefont{Sen}}, \bibnamefont{and}
  \bibinfo{author}{\bibfnamefont{P.~K.} \bibnamefont{Kaw}},
  \bibinfo{journal}{Phys. Rev. Lett.} \textbf{\bibinfo{volume}{101}},
  \bibinfo{pages}{065006} (\bibinfo{year}{2008}{\natexlab{b}}).

\bibitem[{\citenamefont{Pramanik et~al.}(2003)\citenamefont{Pramanik, Veeresha,
  Prasad, Sen, and Kaw}}]{pramanikPLA}
\bibinfo{author}{\bibfnamefont{J.}~\bibnamefont{Pramanik}},
  \bibinfo{author}{\bibfnamefont{B.}~\bibnamefont{Veeresha}},
  \bibinfo{author}{\bibfnamefont{G.}~\bibnamefont{Prasad}},
  \bibinfo{author}{\bibfnamefont{A.}~\bibnamefont{Sen}}, \bibnamefont{and}
  \bibinfo{author}{\bibfnamefont{P.}~\bibnamefont{Kaw}},
  \bibinfo{journal}{Physics Letters A} \textbf{\bibinfo{volume}{312}},
  \bibinfo{pages}{84 } (\bibinfo{year}{2003}), ISSN \bibinfo{issn}{0375-9601}.

\bibitem[{\citenamefont{Khrapak et~al.}(2005)\citenamefont{Khrapak, Ratynskaia,
  Zobnin, Usachev, Yaroshenko, Thoma, Kretschmer, H\"ofner, Morfill, Petrov
  et~al.}}]{khrapakcec}
\bibinfo{author}{\bibfnamefont{S.~A.} \bibnamefont{Khrapak}},
  \bibinfo{author}{\bibfnamefont{S.~V.} \bibnamefont{Ratynskaia}},
  \bibinfo{author}{\bibfnamefont{A.~V.} \bibnamefont{Zobnin}},
  \bibinfo{author}{\bibfnamefont{A.~D.} \bibnamefont{Usachev}},
  \bibinfo{author}{\bibfnamefont{V.~V.} \bibnamefont{Yaroshenko}},
  \bibinfo{author}{\bibfnamefont{M.~H.} \bibnamefont{Thoma}},
  \bibinfo{author}{\bibfnamefont{M.}~\bibnamefont{Kretschmer}},
  \bibinfo{author}{\bibfnamefont{H.}~\bibnamefont{H\"ofner}},
  \bibinfo{author}{\bibfnamefont{G.~E.} \bibnamefont{Morfill}},
  \bibinfo{author}{\bibfnamefont{O.~F.} \bibnamefont{Petrov}},
  \bibnamefont{et~al.}, \bibinfo{journal}{Phys. Rev. E}
  \textbf{\bibinfo{volume}{72}}, \bibinfo{pages}{016406}
  (\bibinfo{year}{2005}),
  \urlprefix\url{https://link.aps.org/doi/10.1103/PhysRevE.72.016406}.

\bibitem[{\citenamefont{Khrapak and Morfill}(2006)}]{khrapakcec2}
\bibinfo{author}{\bibfnamefont{S.~A.} \bibnamefont{Khrapak}} \bibnamefont{and}
  \bibinfo{author}{\bibfnamefont{G.~E.} \bibnamefont{Morfill}},
  \bibinfo{journal}{Physics of Plasmas} \textbf{\bibinfo{volume}{13}},
  \bibinfo{pages}{104506} (\bibinfo{year}{2006}),
  \eprint{https://doi.org/10.1063/1.2359282},
  \urlprefix\url{https://doi.org/10.1063/1.2359282}.

\bibitem[{\citenamefont{Jaiswal
  et~al.}(2016{\natexlab{b}})\citenamefont{Jaiswal, Bandyopadhyay, and
  Sen}}]{surbhi_shock}
\bibinfo{author}{\bibfnamefont{S.}~\bibnamefont{Jaiswal}},
  \bibinfo{author}{\bibfnamefont{P.}~\bibnamefont{Bandyopadhyay}},
  \bibnamefont{and} \bibinfo{author}{\bibfnamefont{A.}~\bibnamefont{Sen}},
  \bibinfo{journal}{Physics of Plasmas} \textbf{\bibinfo{volume}{23}},
  \bibinfo{pages}{083701} (\bibinfo{year}{2016}{\natexlab{b}}),
  \eprint{https://doi.org/10.1063/1.4960032},
  \urlprefix\url{https://doi.org/10.1063/1.4960032}.

\bibitem[{\citenamefont{Epstein}(1924)}]{epstein}
\bibinfo{author}{\bibfnamefont{P.~S.} \bibnamefont{Epstein}},
  \bibinfo{journal}{Phys. Rev.} \textbf{\bibinfo{volume}{23}},
  \bibinfo{pages}{710} (\bibinfo{year}{1924}).

\bibitem[{\citenamefont{Feng et~al.}(2007)\citenamefont{Feng, Goree, and
  Liu}}]{fengrsi}
\bibinfo{author}{\bibfnamefont{Y.}~\bibnamefont{Feng}},
  \bibinfo{author}{\bibfnamefont{J.}~\bibnamefont{Goree}}, \bibnamefont{and}
  \bibinfo{author}{\bibfnamefont{B.}~\bibnamefont{Liu}},
  \bibinfo{journal}{Review of Scientific Instruments}
  \textbf{\bibinfo{volume}{78}}, \bibinfo{pages}{053704}
  (\bibinfo{year}{2007}), \eprint{https://doi.org/10.1063/1.2735920},
  \urlprefix\url{https://doi.org/10.1063/1.2735920}.

\bibitem[{\citenamefont{Jaiswal
  et~al.}(2016{\natexlab{c}})\citenamefont{Jaiswal, Bandyopadhyay, and
  Sen}}]{surabhi_psst}
\bibinfo{author}{\bibfnamefont{S.}~\bibnamefont{Jaiswal}},
  \bibinfo{author}{\bibfnamefont{P.}~\bibnamefont{Bandyopadhyay}},
  \bibnamefont{and} \bibinfo{author}{\bibfnamefont{A.}~\bibnamefont{Sen}},
  \bibinfo{journal}{Plasma Sources Science and Technology}
  \textbf{\bibinfo{volume}{25}}, \bibinfo{pages}{065021}
  (\bibinfo{year}{2016}{\natexlab{c}}),
  \urlprefix\url{https://doi.org/10.1088%2F0963-0252%2F25%2F6%2F065021}.

\bibitem[{\citenamefont{Arora et~al.}(2019{\natexlab{b}})\citenamefont{Arora,
  Bandyopadhyay, Hariprasad, and Sen}}]{garima_second}
\bibinfo{author}{\bibfnamefont{G.}~\bibnamefont{Arora}},
  \bibinfo{author}{\bibfnamefont{P.}~\bibnamefont{Bandyopadhyay}},
  \bibinfo{author}{\bibfnamefont{M.~G.} \bibnamefont{Hariprasad}},
  \bibnamefont{and} \bibinfo{author}{\bibfnamefont{A.}~\bibnamefont{Sen}},
  \bibinfo{journal}{Physics of Plasmas} \textbf{\bibinfo{volume}{26}},
  \bibinfo{pages}{023701} (\bibinfo{year}{2019}{\natexlab{b}}),
  \eprint{https://doi.org/10.1063/1.5078866},
  \urlprefix\url{https://doi.org/10.1063/1.5078866}.

\bibitem[{\citenamefont{Thielicke and Stamhuis}(2014)}]{thielicke2014pivlab}
\bibinfo{author}{\bibfnamefont{W.}~\bibnamefont{Thielicke}} \bibnamefont{and}
  \bibinfo{author}{\bibfnamefont{E.}~\bibnamefont{Stamhuis}},
  \bibinfo{journal}{J. Open Res. Software} \textbf{\bibinfo{volume}{2}},
  \bibinfo{pages}{e30} (\bibinfo{year}{2014}).

\bibitem[{\citenamefont{Annibaldi et~al.}(2007)\citenamefont{Annibaldi, Ivlev,
  Konopka, Ratynskaia, Thomas, Morfill, Lipaev, Molotkov, Petrov, and
  Fortov}}]{annibaldi2007dust}
\bibinfo{author}{\bibfnamefont{S.}~\bibnamefont{Annibaldi}},
  \bibinfo{author}{\bibfnamefont{A.}~\bibnamefont{Ivlev}},
  \bibinfo{author}{\bibfnamefont{U.}~\bibnamefont{Konopka}},
  \bibinfo{author}{\bibfnamefont{S.}~\bibnamefont{Ratynskaia}},
  \bibinfo{author}{\bibfnamefont{H.}~\bibnamefont{Thomas}},
  \bibinfo{author}{\bibfnamefont{G.}~\bibnamefont{Morfill}},
  \bibinfo{author}{\bibfnamefont{A.}~\bibnamefont{Lipaev}},
  \bibinfo{author}{\bibfnamefont{V.}~\bibnamefont{Molotkov}},
  \bibinfo{author}{\bibfnamefont{O.}~\bibnamefont{Petrov}}, \bibnamefont{and}
  \bibinfo{author}{\bibfnamefont{V.}~\bibnamefont{Fortov}},
  \bibinfo{journal}{New Journal of Physics} \textbf{\bibinfo{volume}{9}},
  \bibinfo{pages}{327} (\bibinfo{year}{2007}).

\bibitem[{\citenamefont{Sharma et~al.}(2014)\citenamefont{Sharma, Boruah, and
  Bailung}}]{sharma2014head}
\bibinfo{author}{\bibfnamefont{S. K.}~\bibnamefont{Sharma}},
  \bibinfo{author}{\bibfnamefont{A.}~\bibnamefont{Boruah}}, \bibnamefont{and}
  \bibinfo{author}{\bibfnamefont{H.}~\bibnamefont{Bailung}},
  \bibinfo{journal}{Physical Review E} \textbf{\bibinfo{volume}{89}},
  \bibinfo{pages}{013110} (\bibinfo{year}{2014}).

\bibitem[{\citenamefont{Flanagan and Goree}(2010)}]{flanagan}
\bibinfo{author}{\bibfnamefont{T.~M.} \bibnamefont{Flanagan}} \bibnamefont{and}
  \bibinfo{author}{\bibfnamefont{J.}~\bibnamefont{Goree}},
  \bibinfo{journal}{Physics of Plasmas} \textbf{\bibinfo{volume}{17}},
  \bibinfo{pages}{123702} (\bibinfo{year}{2010}).

\bibitem[{\citenamefont{Merlino et~al.}(2012)\citenamefont{Merlino, Heinrich,
  Kim, and Meyer}}]{merlino2012dusty}
\bibinfo{author}{\bibfnamefont{R.}~\bibnamefont{Merlino}},
  \bibinfo{author}{\bibfnamefont{J.}~\bibnamefont{Heinrich}},
  \bibinfo{author}{\bibfnamefont{S.}~\bibnamefont{Kim}}, \bibnamefont{and}
  \bibinfo{author}{\bibfnamefont{J.}~\bibnamefont{Meyer}},
  \bibinfo{journal}{Plasma Physics and Controlled Fusion}
  \textbf{\bibinfo{volume}{54}}, \bibinfo{pages}{124014}
  (\bibinfo{year}{2012}).

\bibitem[{\citenamefont{Shukla and Eliasson}(2012)}]{shukla2012nonlinear}
\bibinfo{author}{\bibfnamefont{P. K.}~\bibnamefont{Shukla}} \bibnamefont{and}
  \bibinfo{author}{\bibfnamefont{B.}~\bibnamefont{Eliasson}},
  \bibinfo{journal}{Physical Review E} \textbf{\bibinfo{volume}{86}},
  \bibinfo{pages}{046402} (\bibinfo{year}{2012}).

\bibitem[{\citenamefont{Bandyopadhyay et~al.}(2010)\citenamefont{Bandyopadhyay,
  Konopka, Khrapak, Morfill, and Sen}}]{bandyopadhyay2010effect}
\bibinfo{author}{\bibfnamefont{P.}~\bibnamefont{Bandyopadhyay}},
  \bibinfo{author}{\bibfnamefont{U.}~\bibnamefont{Konopka}},
  \bibinfo{author}{\bibfnamefont{S.}~\bibnamefont{Khrapak}},
  \bibinfo{author}{\bibfnamefont{G.}~\bibnamefont{Morfill}}, \bibnamefont{and}
  \bibinfo{author}{\bibfnamefont{A.}~\bibnamefont{Sen}}, \bibinfo{journal}{New
  Journal of Physics} \textbf{\bibinfo{volume}{12}}, \bibinfo{pages}{073002}
  (\bibinfo{year}{2010}).

\bibitem[{\citenamefont{Sen et~al.}(2015)\citenamefont{Sen, Tiwari, Mishra, and
  Kaw}}]{SEN2015429}
\bibinfo{author}{\bibfnamefont{A.}~\bibnamefont{Sen}},
  \bibinfo{author}{\bibfnamefont{S.}~\bibnamefont{Tiwari}},
  \bibinfo{author}{\bibfnamefont{S.}~\bibnamefont{Mishra}}, \bibnamefont{and}
  \bibinfo{author}{\bibfnamefont{P.}~\bibnamefont{Kaw}},
  \bibinfo{journal}{Advances in Space Research} \textbf{\bibinfo{volume}{56}},
  \bibinfo{pages}{429 } (\bibinfo{year}{2015}), ISSN \bibinfo{issn}{0273-1177},
  \bibinfo{note}{advances in Asteroid and Space Debris Science and Technology -
  Part 1},
  \urlprefix\url{http://www.sciencedirect.com/science/article/pii/S0273117715002239}.

\bibitem[{\citenamefont{Arora et~al.}(2018)\citenamefont{Arora, Bandyopadhyay,
  Hariprasad, and Sen}}]{garima_first}
\bibinfo{author}{\bibfnamefont{G.}~\bibnamefont{Arora}},
  \bibinfo{author}{\bibfnamefont{P.}~\bibnamefont{Bandyopadhyay}},
  \bibinfo{author}{\bibfnamefont{M.~G.} \bibnamefont{Hariprasad}},
  \bibnamefont{and} \bibinfo{author}{\bibfnamefont{A.}~\bibnamefont{Sen}},
  \bibinfo{journal}{Physics of Plasmas} \textbf{\bibinfo{volume}{25}},
  \bibinfo{pages}{083711} (\bibinfo{year}{2018}).

\bibitem[{\citenamefont{Truitt and Hartzell}(2020)}]{Truitt2020}
\bibinfo{author}{\bibfnamefont{A.~S.} \bibnamefont{Truitt}} \bibnamefont{and}
  \bibinfo{author}{\bibfnamefont{C.~M.} \bibnamefont{Hartzell}},
  \bibinfo{journal}{Journal of Spacecraft and Rockets}
  \textbf{\bibinfo{volume}{0}}, \bibinfo{pages}{1} (\bibinfo{year}{2020}),
  \eprint{https://doi.org/10.2514/1.A34652},
  \urlprefix\url{https://doi.org/10.2514/1.A34652}.

\end{thebibliography}
\end{document}